\documentclass[twocolumn,prl,showpacs]{revtex4}

\usepackage{graphicx}
\usepackage{dcolumn}
\usepackage{float}
\usepackage{amsmath}

\newcommand{\comment}[1]{}

\begin{document}


\title{Non-BCS behavior of optical properties across the
cuprate phase diagram}
\author{E. Illes$^{1}$}
\author{E.J. Nicol$^{1}$}

\author{J.P. Carbotte$^{2,3}$}

\affiliation{$^1$Guelph-Waterloo Physics Institute and 
Department of Physics, University of Guelph,
Guelph, ON N1G 2W1 Canada} 
\affiliation{$^2$Department of Physics and Astronomy, McMaster
University, Hamilton, ON L8S 4M1 Canada}
\affiliation{$^3$The Canadian
Institute for Advanced Research, Toronto, ON M5G 1Z8 Canada}

\date{\today}

\begin{abstract}
The finite-frequency 
optical properties of the underdoped cuprates, in both the normal
and
superconducting state, display features which go beyond a Fermi liquid
and a BCS 
description.
We provide an understanding of these properties within a
simplified
analytical model, which has been evolved out of the Hubbard model and
ideas based on a resonating valence bond spin liquid. 
We find that: 1) in underdoped samples, the missing area integrals
reveal a second energy
scale due to the pseudogap, not present at optimum or overdoping; 2) 
the
real part of the optical self-energy shows a large sharp peak, that
emerges with the opening of the pseudogap which exists within the
superconducting state and persists in the normal state; 
and 3) the amount of optical
spectral weight which is transferred to the condensate is greatly
reduced by the presence of the 
pseudogap as compared to the Fermi liquid case. These non-BCS
features of the superconducting state
are in good qualitative agreement with a body of experimental work on
different cuprate systems and provide strong evidence from optical
conductivity that they are all a manifestation
of the pseudogap energy scale.
\end{abstract}

\maketitle

Understanding the approach to a Mott insulator from a Fermi liquid
state
is an important question in correlated electron physics. The cuprate
superconductors provide a model system for investigating this
issue. Indeed, they exhibit non-Fermi liquid and non-BCS behavior
throughout an intermediate doping regime between the Mott insulator and the
Fermi
liquid.
In the overdoped
regime, the normal state appears to be rather conventional,
though the superconducting ground state has d-wave symmetry. 
By contrast, the underdoped regime 
exhibits non-Fermi liquid features 
related to the pseudogap formation, which is 
thought to be responsible for
many of its anomalous normal state
properties.\cite{timusk} In addition, in this regime, there is
evidence
that the
pseudogap and superconducting correlations 
compete and both provide
their characteristic imprints on superconducting properties, such as
the
two separate energy scales seen in Raman scattering.\cite{tacon} 
An important bulk probe of materials is finite-frequency optical
conductivity for which a considerable quantity of data has accumulated
about the cuprates as a function of doping and in different systems.
From this it is clear that the optical properties display anomalous
behavior which is non-Fermi liquid and non-BCS-like. This behaviour
has
not been fully 
understood from a theoretical point of view nor has it been
brought together to provide a conclusive statement about the effect
of
a pseudogap energy scale in the superconducting state. In this commnication,
we demonstrate that the non-BCS behavior can be understood as due to a
second competing pseudogap 
energy scale within the superconducting state.
We provide three 
results in comparison with experiment which support this significant
conclusion.

While several approaches to the modeling of the unusual normal state
of the cuprates have been proposed and examined, such 
as d-density waves\cite{chakravarty}
 and phase incoherent preformed pairs above $T_c$\cite{emery},
few have been able to address successfully the superconducting state
or the evolution of properties with doping.
Work on the Hubbard (and related models), thought to be
a good candidate for describing the cuprate phase diagram, has been
largely numerical but recently some progress has been made towards
providing analytic approximations to these works which can facilitate
ease of calculation and provide a deeper understanding of the
underlying physics. In order
to examine the issue of the optical conductivity and its non-BCS
behavior, we have chosen to adopt an analytical approach proposed by
Yang, Rice and Zhang\cite{yrz} which is evolved out of work
that has its basis in the
Hubbard model\cite{konik} and 
the ideas of a resonant valence bond (RVB) spin liquid, proposed
by Anderson\cite{anderson,lee}. 
The coherent part of the electronic Green's
function in this theory encodes the antiferromagnetic Brillouin
zone with pseudogap formation, as the Mott 
transition is approached. We use this formalism to
calculate the in-plane optical conductivity $\sigma(\omega)$ in order
to provide an understanding of the anomalous superconducting 
behavior seen in
quantities
such as the partial optical sum and the optical self-energy.

%
\begin{figure}[t]
  \centerline{\includegraphics[width=3.30 in]{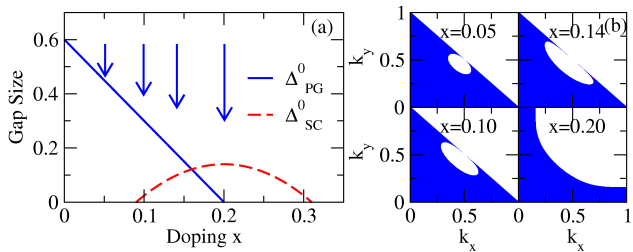}}%
\caption{(color online) 
(a) Maximum value of the pseudogap and superconducting gap in units of
$t_0$ versus
doping ($x$) dependence:\cite{yrz} $\Delta_{pg}^0(x)/2=0.3(1-x/0.2)$ and
$\Delta_{sc}^0(x)/2=0.07[1-82.6(x-0.2)^2]$. (b) Luttinger area in the pseudogap
state for four dopings indicated with arrows in (a). $k_x$ and $k_y$ are in
units of $\pi/a$, with $a$ the lattice spacing.
$G({\bf k},0)>0$ for the shaded region, the area of which satisfies
the Luttinger sum rule.   }
  \label{fig1}
\end{figure}
In the analytical model used here, the coherent part of the electronic 
Green's function takes the form\cite{yrz}
\begin{equation}
G({\bf k},\omega)
=\frac{g_t}{\omega-\epsilon({\bf k})-\Sigma({\bf k},\omega)} ,
\label{eq:G}
\end{equation}
with self-energy
\begin{equation}
\Sigma({\bf k},\omega)=\Sigma_{pg}+\frac{|\Delta_{sc}|^2}{\omega+\epsilon({\bf k})
+\Sigma_{pg}({\bf k},-\omega)} .
\label{eq:Sigma}
\end{equation}
The pseudogap self-energy is given by
$\Sigma_{pg}({\bf k},\omega)=|\Delta_{pg}|^2/[\omega+\epsilon_0({\bf k})]$.
In these formulas $g_t$ is the relative weight of the coherent part of
the Green's function, which we set equal to one. 
There is also an incoherent piece not included here.
The superconducting gap $\Delta_{sc}$ and pseudogap $\Delta_{pg}$ are both assumed
to have d-wave symmetry, i.e. 
$\Delta({\bf k})=\Delta^0(x)[\cos(k_xa)-\cos(k_ya)]/2$,
within the two-dimensional square CuO$_2$ Brillouin zone. 
The doping ($x$)
dependence of their amplitude $\Delta^0(x)$ is 
given in Ref.~\cite{yrz} for a particular case and 
is reproduced here in Fig.~\ref{fig1}a. 
Finally, $\epsilon({\bf k)}$ 
is the electronic dispersion and $\epsilon_0({\bf k})$ the 
first-nearest-neighbor-only version, with $\epsilon_0({\bf k})=0$ 
defining the 
antiferromagnetic Brillouin zone. 
In Fig.~\ref{fig1}b, we show the Luttinger surfaces of the theory
for the parameters of Ref.~\cite{yrz}. The shaded blue areas correspond
to occupied states containing an area proportional to $1-x$ and the white
pockets within the first antiferromagnetic Brillouin zone enclose an area
proportional to $x$.
To compute the in-plane optical conductivity
one needs to know,
in addition to  the Green's function, the Gorkov anomalous amplitude
denoted by $F^\dagger({\bf k},\omega)$ and given by
\begin{equation}
F^\dagger({\bf k},\omega)
=\frac{-\Delta^\dagger_{sc}G({\bf k},\omega)}{\omega+\epsilon({\bf k})+\Sigma_{pg}({\bf k},-\omega)}.
\label{eq:F}
\end{equation}
From these, the conductivity $\sigma(T,\omega)$ follows from the Kubo formula
\begin{eqnarray}
\sigma(T,\nu)&=&\frac{e^2}{2\nu}\sum_{\bf k} v_{\bf k}^2\int^{\infty}_{-\infty}
\frac{d\omega}{2\pi}
[f(\omega)-f(\omega+\nu)]\nonumber\\
\times &&[A({\bf k},\omega)A({\bf k},\omega+\nu)
+B({\bf k},\omega)B({\bf k},\omega+\nu)] ,
\label{eq:cond}
\end{eqnarray}
where  $f(\omega)$ is the Fermi-Dirac distribution function at temperature
$T$, $v_{\bf k}$ is the electron velocity,
 and the spectral functions are given as
$A({\bf k},\omega)=-2{\rm Im} G({\bf k},\omega+i0^+)$
and $B({\bf k},\omega)=-2{\rm Im} F^\dagger({\bf k},\omega+i0^+)$. We present results at $T=0$ 
for the generic band structure and parameters suggested for cuprate systems\cite{yrz}. 
%
%
\begin{figure}[t]
  \centerline{\includegraphics[width=1.65 in]{fig2a.eps}
\vspace*{-5.8 cm}%
\includegraphics[width=1.15 in]{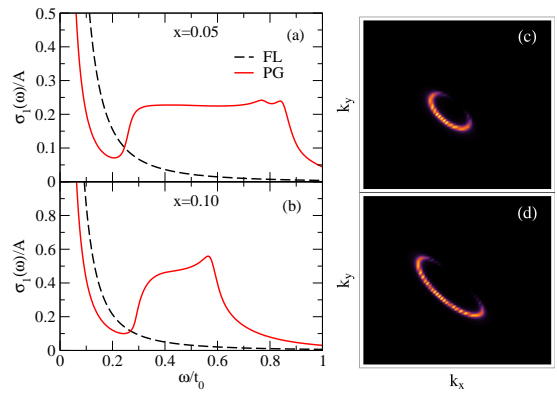}}%
 \vspace*{5.8 cm}%
\caption{(color online) (a) and (b): Optical conductivity curves with 
(solid red curve) and
without (dashed black curve) a pseudogap. 
(c) and (d): Plots of $A({\bf k},\omega=0)$ to show the Fermi arcs, with
$k_x$ and $k_y$ labels normalized as indicated in Fig.~\ref{fig1}.}
  \label{fig2}
\end{figure}
In Fig.~\ref{fig2}, we show $\sigma_1(\omega)/A$ 
as a function 
of photon energy $\omega/t_0$, where $t_0$ sets the energy scale of the first nearest neighbor
hopping parameter. The normalization $A$ has been defined as the area under
the conductivity curve in the normal state for $x=0.2$ (optimally doped).
 Fig.~\ref{fig2}a is for $x=0.05$ (highly underdoped)
and Fig.~\ref{fig2}b is for $x=0.1$, just inside the inner
 edge of the superconducting
dome as shown in the phase diagram of Fig.~\ref{fig1}a.
In Fig.~\ref{fig2}c and d, we plot the corresponding color maps of $A({\bf k},\omega=0)$
which provide information about the Fermi arcs as a function of doping. 
In Fig.~\ref{fig2}a and b, 
the dashed black curves contain no pseudogap and are similar to
a Drude form, while the solid red curves 
contain a finite $\Delta_{pg}$.
In
all our calculations we have replaced the Dirac delta
functions in the spectral densities $A$ and $B$ with Lorentzians of
width $0.01t_0$. Twice this value gives the corresponding transport
width $1/\tau$. The solid red curves, 
in contrast to the dashed curves, deviate strongly from Drude behavior.
The optical spectral weight in the Drude-like distribution centered
at $\omega=0$ is greatly reduced, and some, but not all 
is transferred to
the interband transition 
region.
%
%
\begin{figure}[t]
  \centerline{\includegraphics[width=1.65 in]{fig3a.eps}
  \vspace*{-5.8 cm}%
\includegraphics[width=1.15 in]{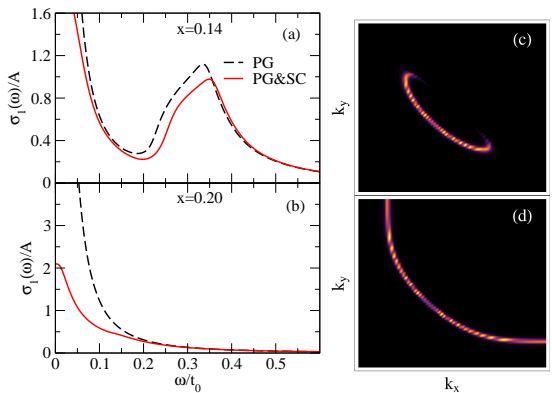}}%
  \vspace*{+5.8 cm}%
\caption{(color online) (a) and (b): Optical conductivity curves 
with (solid red curve) and
without (dashed black curve) a superconducting gap.  The pseudogap is also
present in (a) as can be seen in the phase diagram in Fig.~\ref{fig1}.  
(c) and (d): $A({\bf k},\omega=0)$ showing the Fermi arcs (labels same as
Fig.~\ref{fig2}).}
  \label{fig3}
\end{figure}
The onset of the region to which this weight is transferred, 
is close to $\Delta^0_{pg}$ for $x=0.1$, but falls considerably short of
$\Delta^0_{pg}$ for $x=0.05$. The reason for
this is that the gap value on the Luttinger pockets is much smaller
than $\Delta^0_{pg}$ (see Fig.~\ref{fig2}c). 
For both dopings, the rapid drop
denoting the end of the interband
 region is located at approximately $2\Delta^0_{pg}$. 
 These onset energies can
be traced to peaks in the electronic density of states. 
Not all the optical spectral weight lost in 
the Drude region is recovered in the interband region at higher energies.
We are dealing here only with the coherent part of the
Green's function. In a more complete theory some of the missing weight
could be transferred from coherent to incoherent part. Similar to
our result, a two-component
optical conductivity has also been found in numerical work\cite{kotliar} on
the $t-J$ model.

\begin{figure}[t]
  \centerline{\includegraphics[width=3.30 in]{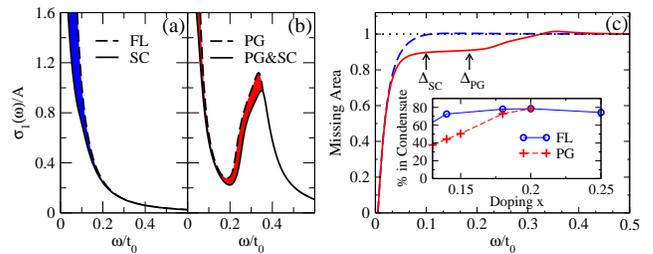}}%
\caption{(color online) 
In (a) and (b) the shaded region represents the missing area, that is 
the spectral weight lost to the condensate 
for $x=0.14$.  The normal state in panel (a) does not
include a pseudogap, whereas in (b) it does.
Panel (c) shows the integral of the missing area $S(\omega)$ normalized to
$S(\infty)$. 
 The inset in (c) shows
the percentage of spectral weight lost to the condensate as a function of
doping for a normal state without (blue circles) and with (red crosses) a
pseudogap.}
  \label{fig4}
\end{figure}

In Fig.~\ref{fig3} we compare the normal state (dashed black curve) with the
corresponding superconducting state (solid red curve) for (a) $x=0.14$
and (b) $x=0.2$. In this latter case,
$\Delta_{pg}=0$
so that the corresponding
Luttinger surface reduces to the usual large Fermi surface of Fermi liquid 
theory (Fig.~\ref{fig3}d), and the real part of the conductivity, $\sigma_1(\omega)$, in 
the superconducting state behaves in a conventional way. 
For an s-wave superconductor there would be a gap of $2\Delta^0_{sc}$ in 
$\sigma_1(\omega)$ with all optical spectral weight 
below this energy going into the condensate and
at higher energies, the normal
state conductivity is recovered. However, here
the gap has d-wave symmetry, and some weight remains
in the Drude region below $2\Delta^0_{sc}$.\cite{ingrid}
Fig.~\ref{fig3}a
is more unusual because the pseudogap, which exists in the
normal state, is responsible for depleting the Drude peak and
transferring weight to the interband region. Adding superconductivity
further depletes the Drude region below $2\Delta^0_{sc}$,
although it does not eliminate it completely. In addition, and this
goes beyond the BCS model, it shifts
the onset of the interband transitions to higher energies. The differences 
between the cases with (frame (a)) and without (frame (b)) pseudogap  are further emphasized in 
Fig.~\ref{fig4}, where we show the real part of the conductivity for $x=0.14$.
The residual scattering rate is $2\eta=0.02t_0$ in both cases.
%
The shaded regions
show the missing area under $\sigma_1(\omega)$
due to the superconducting transition. Frame (a) is conventional d-wave
BCS behavior 
but frame (b) is anomalous. Comparing between (a) and (b),
below $\omega/t_0\approx 0.2$, $\sigma^S_{1}(\omega)$ is not 
affected much by pseudogap formation because the remaining absorption
(Drude) only involves the nodal direction. However, in this energy
region $\sigma^N_{1}(\omega)$ is much larger in (a) than it is in
(b) because in this latter case the antinodal direction is gapped
out with its spectral weight transferred to the interband transitions. Note that
this interband piece contains about 20\% of the 
condensate. This is further elaborated upon 
in (c) where we show results for the missing area partial optical sum
$
S(\omega)=\int_{0^+}^{\omega}[\sigma^N_1(\nu)-\sigma^S_1(\nu)] d\nu.
$
The dashed blue curve is without, while the solid 
red curve is with
a pseudogap. Pseudogap and superconducting energy gap scales are
indicated by arrows. 
The dashed blue curve, 
normalized to $S(\infty)$, rises rapidly to its saturation value
on the scale of the superconducting gap for the particular
residual scattering rate used (a quantity also involved in
determining the rate of initial rise). 
For the solid red curve, this first rise is also
clearly visible but in addition there is a second region of 
 slower rise on an
energy scale set by the pseudogap which involves about 20\% of the condensate. 
This behavior has
been qualitatively
verified in the experimental work of Homes et al.\cite{homesprb}
for  YBa$_2$Cu$_3$O$_{6.95}$ (no pseudogap) and 
YBa$_2$Cu$_3$O$_{6.60}$ (with pseudogap) [see Fig.~7 of Ref.~\cite{homesprb}]. 
A quantative comparison is not possible because
in the analysis
of the experimental data, the normal state at low temperature 
is not available and its value just above $T_c$ is used instead.
Still the pseudogap scale is seen only in  the underdoped case.

In the inset of Fig.~\ref{fig4}c, we show
the fraction of the total available optical spectral weight that goes
into the condensate as a function of doping. For an ordinary s-wave
BCS superconductor this amount would be 100\% in the clean limit,
$1/\tau\ll\Delta^0_{sc}$, as it would also be
for a d-wave gap.
The entire optical spectral weight condenses and the zero
temperature penetration depth depends only on normal state parameters
and not on the value of the gap. For finite $1/\tau$ this 
no longer holds and the amount in the condensate is reduced with the
reduction depending on the relative size of $1/\tau$ to $\Delta^0_{sc}$.
However, when a pseudogap is introduced, the condensate
fraction is further reduced.
This can be traced to the transfer of spectral weight from the Drude
to the region above $\Delta^0_{pg}$ where
it is not as
susceptible to condensation. Note in particular the points
at $x=0.14$. The open circle without pseudogap  is close to 70\%
while the opening of a pseudogap reduces this value to roughly 40\%,
less than half the electrons participate in the condensation.
This is an important result associated with the non-BCS behavior
of this theory and for which there is experimental evidence.\cite{tanner}

%
\begin{figure}[t]
  \centerline{\includegraphics[width=3.30 in]{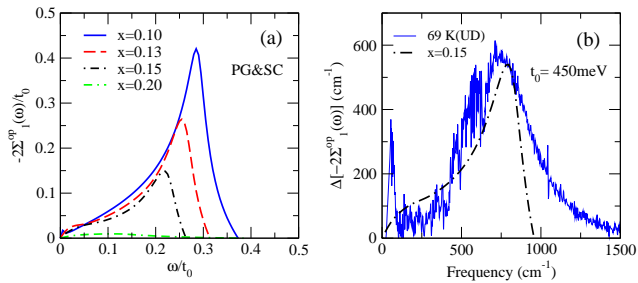}}%
\caption{(color online) (a) Results for $-2\Sigma_1^{op}(\omega)$ vs $\omega$
from the theory (only $x=0.2$ has no pseudogap). (b)
The data of Hwang et al.\cite{bi2212opt}
 for the $\Delta\{-2\Sigma_1^{op}(\omega)\}$ 
vs $\omega$ in cm$^{-1}$ for underdoped Bi2212 with $T_c=67$K  (an incoherent background
has been subtracted out). The black dashed-dotted curve is the
theory for $t_0=450$ meV. Both (a) and (b) show the superconducting
state.} 
  \label{fig5}
\end{figure}
Finally in Fig.~\ref{fig5}, we present our results for the real part of the
optical self-energy $\Sigma^{op}(\omega)$. It has become common in
correlated systems to describe the conductivity $\sigma(\omega)$ in terms
of a generalized Drude form\cite{opqp} with
$\sigma(T,\omega)=(i\omega_p^2/4\pi)/[\omega-2\Sigma^{op}(T,\omega)]$,
where $\omega_p$ is the plasma frequency.
For the 
identification of the pseudogap in in-plane optics it was recently found
that $\Sigma^{op}_1(T,\omega)$ is a more useful
 quantity\cite{hwanghat} than is the conductivity
itself, and it shows a prominent ``hat''-like structure seen above
a large inelastic background.  
Prominent peaks in $\Sigma^{op}_1(T,\omega)$ can arise from either
pseudogap or superconducting gap formation\cite{hwang05,bi2212opt}
when the inelastic scattering, described by the incoherent part of the 
Green's function, involves sharp boson structure on the same energy
scale.  In the present work such structures arise even though
inelastic
scattering is not accounted for. While it is not clear exactly how
to separate coherent and incoherent contributions,
 Hwang et al.\cite{bi2212opt} have removed a
background from their data presented in their Fig.~19 for the
Bi$_2$Sr$_2$CaCu$_2$O$_{8+\delta}$ (Bi2212) series. 
The results of our calculations,
 which
are shown in Fig.~\ref{fig5}a for various values of doping $x$,
are best compared with these data for one of the most underdoped samples
studied with $T_c=67$K, reproduced in Fig.~\ref{fig5}b (solid blue curve).
Pseudogap effects are large in this case and dominate the peak formation,
which is also hardly affected by superconductivity in both experiment and theory. Taking $t_0=450$ meV
and $x=0.15$, the black dot-dashed curve is reproduced in Fig.~\ref{fig5}b.
The agreement with data is reasonable particularly as we have not
used parameters specific to Bi2212 but instead stayed with those for 
Ca$_{2-x}$Na$_x$CuO$_2$Cl$_2$ given in Ref.~\cite{yrz}. 
The subtraction of the incoherent contribution to 
$\Sigma_1^{op}(T,\omega)$ may also be responsible for some of the discrepancies
seen beyond $\sim 900$~cm$^{-1}$. As stated the peaks of Fig.~\ref{fig5}a 
are greatly reduced with increasing doping,
a trend confirmed in the normal state experimental data.

In summary,
we have been able to explain 
three major anomalous experimental observations from finite-frequency optical
conductivity measurements on several different cuprate systems
where non-BCS behavior is exhibited as a function
of doping. 
Such non-BCS behavior is understood in our work to result from
a pseudogap-type energy scale present in the normal state which also
manifests itself within the superconducting state. 
Comparison of our theoretical results with optical
data in the Bi2212 series reveals good qualitative agreement for the
evolution with doping of both position in energy and size
of the normal state ``hat''-like structures seen in the real part of 
the optical self-energy. 
In addition, the 
partial optical sum reveals that a second energy scale related to the
pseudogap exists in underdoped samples and is not present
in optimally or overdoped samples, as seen in the
data of Homes et al\cite{homesprb} for YBa$_2$Cu$_3$O$_{7-\delta}$.
Another significant result of our analysis is that the opening of a pseudogap
greatly reduces the percentage of carriers that condense in the
superconducting state as compared to BCS.
We stress that optical conductivity is a
bulk probe applicable to many cuprate systems, 
as opposed to surface probes such as scanning
tunneling and angle-resolved photoemission 
spectroscopies  which are
mainly limited to the Bi2212 system.
Consequently our analysis provides strong
support that the observed non-BCS behaviors seen in
several cuprate systems and their doping dependence can
be described via a theory, such as we have used here,
based in the Hubbard model and ideas of a RVB spin liquid. 

We thank J. Hwang for supplying us with his data
and I. Vekhter for his comments.
This work has been supported by the Natural Sciences and
Engineering Research Council of Canada (NSERC) and the Canadian Institute
for Advanced Research (CIFAR).

\end{document}